\documentclass{elsart}
\usepackage{epsfig}
\newcommand{\beq}{\begin{equation}}
\newcommand{\eeq}{\end{equation}}

\begin{document}
\begin{frontmatter}
\title{Improved theoretical prediction for the $2s$ hyperfine
interval in helium ion}
\author[vniim,mpq]{Savely G. Karshenboim}
\ead{sek@mpq.mpg.de} \and
\author[gao,vniim]{Vladimir G. Ivanov}
\address[vniim]{D. I. Mendeleev Institute for Metrology (VNIIM),
St. Petersburg 190005, Russia}
\address[mpq]{Max-Planck-Institut f\"ur Quantenoptik, 85748 Garching,
Germany}
\address[gao]{Pulkovo Observatory, 196140 St. Petersburg, Russia}

\begin{abstract}
We consider the uncertainty of theoretical calculations for a specific
difference of the hyperfine intervals in the $1s$ and $2s$ states
in a light hydrogen-like atom. For a number of crucial radiative
corrections the result for hydrogen atom and helium ion appears as
an extrapolation of the numerical data from medium to low  $Z$.
An approach to a plausible
estimation of the uncertainty is
suggested using the example of the difference $D_{21}=
8E_{\rm hfs}(2s)-E_{\rm hfs}(1s)$.
\end{abstract}

\end{frontmatter}


Studies of the hyperfine structure in light hydrogen-like atoms are
of interest because of a possibility for a precision test of the
bound state Quantum Electrodynamics (QED). However, in spite of a
record accuracy achieved in experiments on the hyperfine interval
in the ground state of hydrogen, deuterium, tritium and helium-3
ion (see, e.g., \cite{ramsey}), we should acknowledge that such a
test cannot be really successful until effects of the nuclear
structure are known accurately enough.

In fact, the bound state QED effects are lower than those of the
nuclear structure. Obviously, the latter cannot be known well
enough. However, there is a hopeful opportunity to perform such a
test if we consider data of two measurements: for hyperfine
intervals in the ground and metastable states (i.e. for the $1s$
and the $2s$ states).

Combining the difference
\begin{equation}
D_{21}= 8E_{\rm hfs}(2s)-E_{\rm hfs}(1s)\,,
\end{equation}
we take advantage of a substantial cancellation of various
contributions caused by short distance effects, since
\begin{equation}
\frac{\vert\psi_{1s}(0)\vert^2}{\vert\psi_{2s}(0)\vert^2}=8\;,
\end{equation}
where $\psi_{ns}(0)$ is a value of the Schr\"odinger-Coulomb wave
function at origin for the $ns$ state.

The nuclear structure is such a short-distance effect and its
leading contribution is proportional to $\vert\psi(0)\vert^2$.
That is the crucial feature of the difference $D_{21}$ that a
complete cancellation of the leading nuclear structure term takes
place. Meanwhile the higher order nuclear effects are suppressed
by the factor of $(Z\alpha)^2$ and are well under control (see
\cite{sgkpsas,d21} for detail).


Theoretical contributions are conventionally presented in terms of the
so-called Fermi energy, which determines the hyperfine interval
for the ground state. For the nuclear spin 1/2 it is of the form
\begin{equation}
E_F =
  {8 \over 3} \,Z^3 \alpha^4 \,m \, {\mu \over \mu_B}\;,
\end{equation}
which involves the nuclear magnetic moment $\mu$ and the Bohr
magneton $\mu_B$.

Throughout the paper we apply units in which $\hbar=c=1$; $m$ is
the electron mass; $M$ is the nuclear mass; $Z$ is the nuclear
change; and while all equations are presented for the energy
($\Delta E$), the numerical results are for the related frequency
($\Delta E/h$); all fractional values are in units of $E_F$.

The leading contribution to the difference $D_{21}$ is
\begin{equation}
D_{21}^{(0)} = {5 \over 8}\,(Z\alpha)^2 \,E_F
\end{equation}
and the known corrections include higher order QED terms (of the
third and fourth order in the expansion over $\alpha$, $Z\alpha$
and $m/M$) and higher order nuclear effects. We mainly follow
\cite{d21} after correcting a misprint  in the expression for the nuclear
magnetic moment of the helion (the nucleus of the helium-3 ion).
The correction slightly shifts the result of \cite{d21}
($D_{21}^{\rm theor}({}^3{\rm He}^+)=-1\,190.067(63)\;{\rm kHz}$)
to $D_{21}^{\rm theor}({}^3{\rm He}^+)=-1\,190.083(63)\;{\rm kHz}$
(see,e.g., \cite{pos}). Other corrections, but one, have a
marginal effect and will be considered elsewhere.


Here, we consider the most important theoretical issue which
affects our early predictions for $Z=1,2$. Before discussing it
let us remind that a substantial part of the theoretical
uncertainty in calculations of \cite{d21} came from the one-loop
self energy contribution  estimated according to
Ref.~\cite{yero}. It is of approximately the same value as
contributions to the error budget in \cite{d21} from the two-loop
corrections and the higher-order nuclear-size effects. A
conclusion of the present investigation is that it was originally
underestimated  \cite{yero} and actually it is
approximately four times larger and thus gives the dominant
effect to the uncertainty.

The one-loop self-energy term can be presented in the form
\begin{equation}\label{e:one}
D_{21}^{\rm SE}(Z) = \frac{\alpha}{\pi} \, (Z\alpha)
  ^2 \,F_{\rm SE}(Z)\,E_F\;,
\end{equation}
where
\begin{equation}
F_{\rm SE}(Z) =  \left[
    a_{21} \ln \frac{1}{Z\alpha}
    + a_{20}
    +  (Z\alpha) C_{\rm SE}(Z)
  \right]
 \,.
\end{equation}

The coefficients $ a_{21}$ and $ a_{21}$ are known
\cite{zwanziger,mohr} (see, e.g., \cite{d21} for detail)
\begin{eqnarray}\label{e:num}
a_{21}&=&7-{16 \over 3} \,\ln2=3.303\,22\dots\;,\nonumber\\
a_{20}&=& -5.221\,23\dots
\end{eqnarray}
and evaluation of $C_{\rm SE}(Z)$ is a purpose of our study.


A perturbative evaluation of this coefficient has not yet been
done. The results of \cite{yero} and this paper were obtained by
fitting the numerical data obtained in \cite{yero} for a medium
$Z$ region, while the value of interest is related to $Z=1,2$ .
The calculation itself is a complicated problem, in fact much more
complicated than the extrapolation. However, it is of a minor
importance if the extrapolation is not performed properly.

Some time ago we suggested an appropriate approach
described in part in \cite{interp}. In fact, the authors of
\cite{yero} stated that they applied our procedure and reached
\begin{eqnarray}\label{e:num:yero}
 C_{\rm SE}(1)&=&2.07 \pm 0.25\;,\nonumber\\
 C_{\rm SE}(2)&=&2.01 \pm 0.19\;.
\end{eqnarray}
The idea of the procedure, which we present here in more detail,
is based on acknowledging a few typical features of any
extrapolation for the self-energy contribution.
\begin{itemize}
\item The series contain numerous logarithmic contributions.
\item The data are available for medium $Z$ (say from 10 to 30).
\item The slowly-changing logarithmic part ($\ln(Z)$) can be
hardly separated from the nonlogarithmic terms, which may include
$\ln(\alpha)$.
\item Because of that the strategy of the fitting it to exclude
any logarithmic dependence on the first stage and perform a
polynomial fit, e.g. $C(Z)=A_0 + A_1Z + A_2Z^2$. Since each $Z$ is
accompanied with $\alpha$, the coefficients decrease with their
order $A_0\gg A_1 \gg A_2$.
\item The uncertainty of the fit is determined by stability of the
results against a perturbation by logarithmic terms, e.g. by a
difference with the results from fits $C(Z)=A_0 + A_1Z + A_2Z^2 +
\Delta C_{\rm log}$, where $\Delta C_{\rm log}$ are various
possible logarithmic contributions with coefficient of the natural
value.
\end{itemize}

However, a crucial part of the procedure is estimation of values
of coefficients for the logarithmic terms which is explained in
detail below while applying to the one-loop self energy
contribution to $D_{21}$.


The most general expression, which includes all contributions
 to $C_{\rm SE}(Z)$ up to order $(Z\alpha)^3$ is of the form
\begin{eqnarray}\label{e:coeff}
  C_{\rm SE}(Z) = a_{30}  &+& (Z\alpha) \left( a_{42} \ln^2 \frac{1}{Z\alpha}
  + a_{41} \ln \frac{1}{Z\alpha} + a_{40} \right)\nonumber\\
  &+& (Z\alpha)^2 \left( a_{51} \ln \frac{1}{Z\alpha} + a_{50} \right)
  + \dots
  \,.
\end{eqnarray}
The inclusion of higher order terms is not necessary here, since
it will not improve accuracy of the fitting over $Z\leq 30$.

\begin{figure*}[tbp]
\begin{center}
\includegraphics[width=0.7\textwidth]{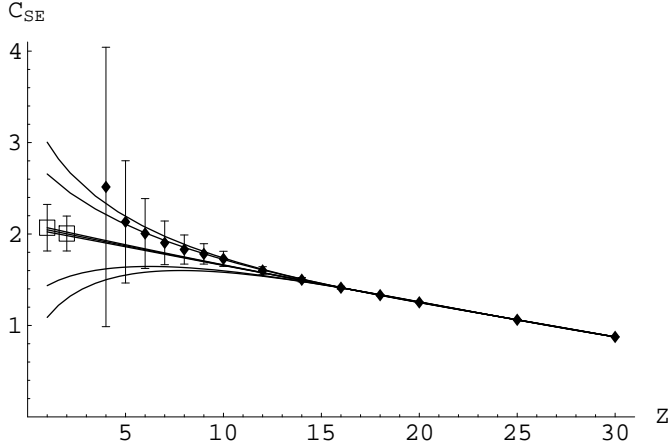}
\end{center}
\caption{Parabolic extrapolation of numerical data taken from
\cite{yero} and its perturbation by various logarithmic functions.
The open squares are the results of the original extrapolation in
\cite{yero}.} \label{f:fit}
\end{figure*}

The central value of the extrapolation according to our procedure
can be obtained by setting $a_{42}=a_{41}=a_{51}=0$. That is
related to a parabolic fit (see Fig.~\ref{f:fit}) and on this
issue we agree with results from \cite{yero}. However, we strongly
disagree on the uncertainty of the result, i.e. we
estimate a possible shift of such a result
in a different way  by using the logarithmic
perturbation.


A crucial question while estimating such a shift properly is related to
the natural value of the coefficients.
It is worth noting that odd and even terms of the series have
different structure and magnitude of the coefficients. With this
observation we estimate
\begin{eqnarray}\label{e:num:fit}
a_{42}&=&\pm a_{21} \simeq \pm 3.30\;,\nonumber\\
a_{41}&=&\pm {\rm max}\Bigl(\pi a_{21}, a_{20}\Bigr) \simeq \pm 10.4\nonumber\\
a_{51}&=&\pm a_{30} \approx \pm 2\;,
\end{eqnarray}
where we apply a preliminary value of $a_{30}$ from the parabolic
extrapolation, which is sufficient for this purpose. Here a factor
of $\pi$ appears because that is a characteristic value of the
constant beyond the logarithm. The natural values of the unknown
coefficients are pretty large and their underestimation should
lead to a serious overestimation of the accuracy of the
extrapolation $C(Z)$ to $Z=1,2$ (cf.~\cite{yero}).


While the introduced higher order corrections are  negligible for
$Z=1,2$, they are not small for $Z=10-30$, i.e. in the range of
the input data for the extrapolation. The data \cite{yero} for $Z$
below 10 look correlated and uncertain and we do not use them for
the fitting. The same was done in \cite{yero}.

Results of our extrapolation are summarized in Fig.~\ref{f:fit}
and Table~\ref{t:fits}. A scatter of the extrapolation results
around the parabolic values allows us to estimate the uncertainty and
our final results
\begin{eqnarray}\label{e:num:us}
 C_{\rm SE}(1)&=&2.1 \pm 1.0\;,\nonumber\\
 C_{\rm SE}(2)&=&2.0\pm0.7
\end{eqnarray}
are approximately four times less accurate than in
(\ref{e:num:yero}). Since we consider logarithmic coefficients of
order of unity in the natural units, any more accurate result may
be achieved only on base of additional information and we consider
the uncertainty in \cite{yero} as unappropriate.

We have to comment briefly the stability of the fits. The
parabolic coefficients $a_{42}$ and $a_{41}$ look unstable,
however, one has to understand, that they are not related to any
`true' coefficient in (\ref{e:coeff}). They are `effective'
coefficients. The logarithmic terms $\ln(Z)$ slowly depend on $Z$
and effectively the effective non-logarithmic coefficients include
$\ln(Z_0)$, where $Z_0\simeq 20$ is kind of a medium $Z$ value. If
we like to see stability of the non-logarithmic coefficients, we
have to perturb the parabolic fit by $\ln(Z)-\ln(Z_0)$. Such a fit
would be more stable, however, mathematically the procedure is
equivalent to what we have done. The actual stability should be
seen not through values of auxiliary fitting parameters, but as a
distribution of the results of the extrapolation $C_{\rm SE}(1)$
and $C_{\rm SE}(2)$.

\begin{table}[tbp]
\caption{Parameters of the parabolic extrapolation (0) and its
perturbation by the logarithmic functions (1--6). The logarithmic
parameters are introduced as a fixed perturbation, while the
non-logarithmic terms are found via least-square minimization. The
uncertainty shown is a pure statistical one.} \label{t:fits}
\bigskip
\begin{center}
\begin{tabular}{|l|r|r|r|r|r|r|r|}
\hline
{Coeff}&{0}&{1}&{2}&{3}&{4}&{5}&{6}\\
\hline
$a_{30}$             &2.09(6)    &3.60(6)    & 0.58(8)   &   2.94(6)    &   1.24(8)   &    2.12(6)  &   2.06(6)  \\
$a_{42}$             &0          &3.3          &-3.3         &0             &0            &0            &0           \\
$a_{41}$             &0          &0          &0          &10.38          &-10.38        &0            &0           \\
$a_{40}$             &6.1(6)     &-1.98(6)   &-10.2(9)   &   12.5(6)    &   -24.7(8)  &    -6.8(6)  &   -5.4(7)  \\
$a_{51}$             &0          &0          &0          &0             &0            &2            &-2          \\
$a_{50}$             &2.4(1.8)   &-13.1(1.7) & 17.9(2.4) &   -28.4(1.7) &   33.2(2.3) &    1.8(1.7) &   3.0(7)   \\
$C_{\rm SE}(1)$&2.05(5)    & 3.00(5)   & 1.09(7)   & 2.66(5)      & 1.43(7)     & 2.08(5)     & 2.01(5)    \\
$C_{\rm SE}(2)$&2.00(5)    & 2.71(5)   & 1.29(7)   & 2.48(5)      & 1.53(6)     & 2.03(5)     & 1.97(5)    \\
\hline
\end{tabular}
\end{center}
\end{table}


We have calculated above the one-loop self-energy contribution for
hydrogen and deuterium ($Z=1$) and heluim-ion ($Z=2$). The latter
is of most interest since it is a more sensitive test of bound
state QED (see \cite{d21} for detail).

Finally, we obtain for the helium ion
\begin{equation}\label{e:d21:th}
D_{21}^{\rm theor}({}^3{\rm He}^+) = -1\,190.08(15)\;{\rm kHz}\;,
\end{equation}
while the most accurate experimental result is
\cite{exphe1s,prior}
\begin{equation}\label{e:d21:e}
D_{21}^{\rm exp}({}^3{\rm He}^+) = -1\,189.979(71)\;{\rm kHz}\;.
\end{equation}

\begin{figure*}[tbp]
\begin{center}
\includegraphics[width=0.6\textwidth]{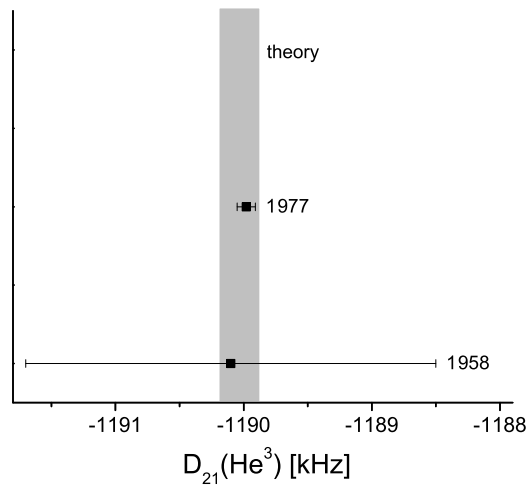}
\end{center}
\caption{The $D_{21}$ difference for the helium ion ${}^3$He$^+$:
theory versus experiment.} \label{f:he}
\end{figure*}

A comparison of theory and experiment is summarized in
Fig.~\ref{f:he}, where the experimental data are labelled with the
date of measurements of the hyperfine interval of the $2s$ state:
1958 \cite{exphe2s} and 1977 \cite{prior}. Meanwhile the $1s$
hyperfine interval was obtained experimentally in \cite{exphe1s}.

Theory is in  perfect agreement with experiment. The most
important consequence of our conservative estimation of the
theoretical uncertainty is the fact that the experimental
result (\ref{e:d21:e}) is now twice more accurate than that of the
theoretical prediction (\ref{e:d21:th}), while previously the
relation was opposite.

The procedure presented here has allowed to estimate uncertainty
of the extrapolation of the numerical data properly for $D_{21}$.
Such an extrapolation occurs for a number of problems in QED
theory for hydrogen-like atoms, and a plausible estimation of the
accuracy is a crucial issue for the comparison of precision
theoretical and experimental results. We hope that an application of
our procedure will be helpful for other problems.

The hydrogen and deuterium data will be discussed elsewhere
together with experimental progress in the field.


The authors are grateful to S. I. Eidelman for useful discussions. 
A part of this work was done during a visit of VGI to Garching and
he is grateful to MPQ for their hospitality. This work was
supported in part by the RFBR grants 03-02-04029 and 03-02-16843
and DFG grant GZ 436 RUS 113/769/0-1.


\end{document}